\documentclass[11pt]{article}

\usepackage[preprint]{acl}

\usepackage{times}
\usepackage{latexsym}
\usepackage{booktabs} 
\usepackage{amsmath}
\usepackage{amssymb}
\usepackage{mathtools}
\usepackage{amsthm}
\usepackage{microtype}
\usepackage{hyperref}
\usepackage{url}
\usepackage{booktabs}
\usepackage{microtype}

\usepackage{subcaption}
\usepackage{booktabs} 
\usepackage{booktabs} 
\usepackage{multirow} 
\usepackage{booktabs}   
\usepackage{makecell}   
\usepackage{multirow}   
\usepackage[table]{xcolor} 
\usepackage{pifont}
\usepackage[most]{tcolorbox}
\usepackage{cleveref} 
\usepackage{hyperref}

\usepackage{amsmath}
\usepackage{amssymb}
\usepackage{mathtools}
\usepackage{amsthm}
\usepackage[T1]{fontenc}

\usepackage[utf8]{inputenc}

\usepackage{microtype}

\usepackage{graphicx}
\usepackage{xcolor}
\usepackage{tcolorbox}
\usepackage{listings}
\usepackage{subcaption}

\usepackage{booktabs}

\tcbuselibrary{breakable, skins}

\usepackage[most]{tcolorbox}
\usepackage{xcolor}
\usepackage{enumitem}
\usepackage{hyperref}
\usepackage{cleveref}
\usepackage{sourcesanspro}
\usepackage[varqu,varl]{inconsolata}
\usepackage[table]{xcolor}
\definecolor{promptbg}{HTML}{F8FAFC}
\definecolor{promptborder}{HTML}{CBD5E1}
\definecolor{promptheader}{HTML}{0F172A}
\definecolor{promptlabel}{HTML}{1E293B}
\definecolor{prompttext}{HTML}{334155}
\definecolor{promptplaceholderbg}{HTML}{E2E8F0}
\definecolor{promptplaceholder}{HTML}{0F172A}

\newtcolorbox[auto counter]{promptbox}[2][]{
    enhanced,
    breakable,
    colback=promptbg,
    colframe=promptborder,
    coltitle=white,
    colbacktitle=promptheader,
    title=\textbf{Prompt~\thetcbcounter: #2},
    fonttitle=\small\sffamily,
    fontupper=\small\sffamily\color{prompttext},
    boxrule=0.7pt,
    arc=2mm,
    left=3mm,
    right=3mm,
    top=2mm,
    bottom=2mm,
    toptitle=1mm,
    bottomtitle=1mm,
    titlerule=0pt,
    before upper={
        \setlength{\parindent}{0pt}
        \setlength{\parskip}{4pt}
        \linespread{1.03}\selectfont
    },
    #1
}

\newcommand{\promptfield}[1]{%
    \vspace{2pt}
    {\color{promptlabel}\textbf{#1}}%
}

\newcommand{\placeholder}[1]{%
    \tcbox[
        on line,
        colback=promptplaceholderbg,
        colframe=promptplaceholderbg,
        arc=1mm,
        boxrule=0pt,
        left=1mm,
        right=1mm,
        top=0.3mm,
        bottom=0.3mm
    ]{\texttt{\color{promptplaceholder}\{#1\}}}%
}

\crefname{tcbcounter}{Prompt}{Prompts}

\title{HMARS: A Hierarchical Multi-Agent Memory System for Long-Context Reasoning}

\author{
 \textbf{Zeju Li\textsuperscript{1}},
 \textbf{Ziyang Zheng\textsuperscript{1}},
 \textbf{Yizhou Zhou\textsuperscript{2}\textdagger},
 \textbf{Qiang Xu\textsuperscript{1}\textdagger},
\\
 \textsuperscript{1}The Chinese University of Hong Kong,
 \textsuperscript{2}Independent Researcher
\\
 \small{
   {zjli24@cse.cuhk.edu.hk, zyzheng23@cse.cuhk.edu.hk, zyz0205@hotmail.com, qxu@cse.cuhk.edu.hk}
 }
\\
\small{\textdagger Corresponding Author}
}
\usepackage{xcolor}

\begin{document}
\maketitle
\begin{abstract}
Long-context reasoning requires models to access, retrieve, and integrate evidence scattered across documents, dialogues, and accumulated interaction histories. 
Standard retrieval-augmented generation reduces this problem to top-$K$ chunk retrieval, but such passive access can discard relevant evidence before reasoning begins, especially when relevance depends on broader context. 
We propose \textsc{HMARS}, a hierarchical multi-agent memory system that treats long contexts as managed memory rather than a flat retrieval corpus. 
Sub-agents maintain grounded access to bounded memory regions, mid-agents manage regional context and provide query-specific coordination, and a frontier model performs final reasoning over retrieved evidence pages. 
To evaluate this view, we construct two diagnostic benchmarks targeting evidence breadth and context-dependent relevance. 
Across long-document and multi-turn memory tasks, \textsc{HMARS} achieves the best overall performance against retrieval, reranking, full-context, graph-based, and agentic long-context baselines. 
Evidence coverage analysis further shows that its gains come from retrieving the required supporting evidence more completely, rather than merely changing the final answer prompt.
\end{abstract}

\section{Introduction}
\label{sec:introduction}

Large language models \citep{yang2025qwen3technicalreport, comanici2025gemini, achiam2023gpt, singh2026openaigpt5card} are increasingly expected to reason over long documents, multi-turn dialogues, and accumulated interaction histories. 
However, simply expanding context windows does not solve long-context reasoning: inference remains expensive, attention over long inputs is unreliable, and relevant evidence is often buried among distractors. 
Retrieval-augmented generation (RAG) \citep{jiang2024longrag, lewis2020retrieval} remains the default solution \citep{bai2024longbench,bai2025longbenchv2}: it chunks documents, embeds chunks independently, retrieves the top-$K$, and passes them to a generator. 
This passive interface is efficient, but assumes relevance is local and statically measurable. 
The assumption breaks when evidence is scattered across distant regions or becomes relevant only under broader context---for example, an early ``\$2,500 budget'' may be essential to a later question about ``housing constraints'' despite low lexical overlap. 
RAG stores context; it does not provide a memory system for managing, accessing, and reasoning over evidence.

\begin{figure}[t]
    \centering
    \includegraphics[width=1.0\linewidth]{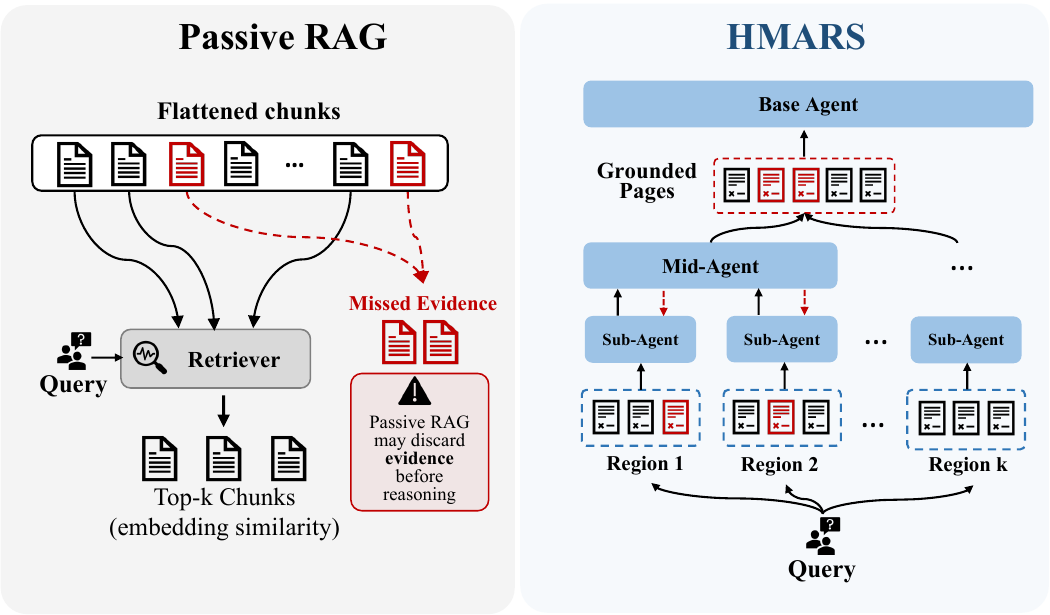}
    \vspace{-20pt}
    \caption{Passive RAG chunks documents, embeds chunks independently, and retrieves top-$K$ evidence, but may miss context-dependent evidence. \textsc{HMARS} treats long inputs as managed memory, where sub-agents assess local regions, mid-agents coordinate shared context, and a frontier model performs final reasoning.}
    \vspace{-20pt}
    \label{fig:comparsion}
\end{figure}

Recent work has begun to move beyond passive retrieval, but still falls short of a memory system for long-context reasoning. 
Memory-augmented agents \citep{chhikara2025mem0,xu2026mem,zhou2025mem1,zhang2026himem,sun2026h} mainly study what to store, update, compress, or forget over long-horizon interactions. 
Long-context retrieval and reading methods \citep{jiang2024longrag,sarthi2024raptor,lee2024readagent} improve access through larger chunks, summary hierarchies, or page lookup, but remain passive or centralized: some module still decides which parts of the input are allowed to reach the reasoner. 
What is missing is a memory system that organizes long inputs into local regions, lets each region participate in query-conditioned access, retrieves grounded evidence, and leaves final reasoning to a stronger model.

We propose \textsc{HMARS}, a Hierarchical Multi-Agent Memory System for long-context reasoning. 
\textsc{HMARS} treats long documents and interaction histories as managed memory rather than a flat chunk corpus. 
Lightweight sub-agents retain grounded access to bounded memory regions, mid-agents manage shared regional context and provide query-specific coordination, and a frontier base model performs final reasoning over retrieved evidence. 
Unlike top-$K$ retrieval, \textsc{HMARS} does not pre-filter away document regions: every sub-agent performs query-conditioned assessment over its local memory, while mid-agent guidance helps retrieve focused, grounded evidence, shown in Figure \ref{fig:comparsion}.

To evaluate this memory-system view, we construct two diagnostic benchmarks that stress evidence breadth and context-dependent relevance rather than short sequential hops. 
The first is derived from LongBench-v2 \citep{bai2025longbenchv2}, with questions filtered or generated to require evidence across distant chunks and low RAG recall; the second is a cumulative interaction benchmark inspired by RealMem \citep{bian2026realmem}, where key facts are scattered across multi-turn interactions and become relevant only under later context. 
Across these settings, \textsc{HMARS} improves most where passive top-$K$ retrieval and centralized memory access fail, supporting our core claim: long-context reasoning should be framed not only as retrieving more chunks, but as managing, accessing, retrieving, and reasoning over memory.


Our contributions are threefold. 
First, we build harder diagnostic benchmarks for long-context reasoning and multi-turn interaction memory, where questions require broader evidence coverage across distant chunks, regions, or turns. 
Second, we frame these challenges as failures of passive chunk-based memory access, especially under evidence breadth and context-dependent relevance. 
Third, we introduce \textsc{HMARS}, a hierarchical multi-agent memory system that separates local memory access, mid-level coordination, grounded evidence retrieval, and frontier-model reasoning.

\section{Related Work}

\subsection{Long-Context Retrieval and Hierarchical Reading}

Retrieval-augmented generation remains the dominant interface for long-context QA \citep{singh2026agenticretrievalaugmentedgenerationsurvey,jin2025longrefiner,gupta2024comprehensive}. 
Recent work improves this pipeline through adaptive retrieval and critique \citep{asai2024self}, retrieval correction, larger retrieval units \citep{jiang2024longrag}, hierarchical summary trees \citep{sarthi2024raptor}, or gist memory with page lookup \citep{lee2024readagent}. 
These methods improve access to long context, but still rely on passive or centralized selection: a retriever selects chunks, a summary tree is searched, or a single reader controls lookup. 
\textsc{HMARS} instead treats long context as an agent-managed memory hierarchy, where local regions participate in query-conditioned assessment and return grounded pages without early top-$K$ exclusion.

\subsection{Multi-Agent Long-Context Processing}

Multi-agent frameworks have also been explored for long-context reasoning \citep{wang2025mirix,jiang2025cocoa,zhao2024longagent}. 
Chain-of-Agents \citep{zhang2024coa}, for example, assigns chunks to sequential worker agents that pass compressed communication units before a manager produces the final answer. 
These methods show that agent collaboration can reduce input bottlenecks, but they still treat long-context processing primarily as \emph{sequential reading}: each query triggers a pass over the input, and information is propagated through compressed messages. 
\textsc{HMARS} instead treats the input as a persistent memory system: sub-agents retain grounded page access to local regions, mid-agents provide shared regional context and query-specific coordination, and the base model reasons over retrieved evidence.

\subsection{Memory-Augmented LLM Agents}

A growing line of work studies explicit memory systems for LLM agents. 
Mem0 \citep{chhikara2025mem0} extracts and retrieves salient conversational memories; A-Mem \citep{xu2026mem} organizes memories as dynamically linked notes; MEM1 \citep{zhou2025mem1} jointly learns memory consolidation and reasoning; HiMem \citep{zhang2026himem} builds hierarchical Episode and Note memories; and reflective or learned memory managers study how agents should update, compress, or delete memories over time \citep{tan2025prospect,yan2025memory,yu2025memagent}. 
While these works establish memory as a central substrate for LLM agents, they mainly focus on the lifecycle of persistent memories: what to store, update, consolidate, or forget. 
\textsc{HMARS} addresses the complementary problem of accessing a given long context as memory, so that local regions can be queried, grounded evidence can be retrieved, and no region is discarded before query-conditioned assessment.

\section{Method}
\label{sec:method}

We present \textsc{HMARS}, a Hierarchical Multi-Agent Memory System for long-context reasoning over documents and interaction histories. 
Rather than treating a long input as a flat corpus of retrievable chunks, \textsc{HMARS} organizes it as a hierarchy of local memory regions. 
Sub-agents retain grounded page-level access to bounded regions, mid-agents coordinate these local memories through shared regional context and query-specific guidance, and a frontier base model performs final reasoning over the retrieved evidence pages.

\subsection{Problem Setup}
\label{sec:problem}

Given a long document or interaction history $\mathcal{D}$ and a query $q$, the goal is to produce an answer $a$ using a compact evidence context $\mathcal{C}(q) \subset \mathcal{D}$. 
We focus on settings where the answer cannot be reliably recovered from a small number of independently retrieved chunks: evidence may be scattered across distant document regions, or a passage may become relevant only under information stated elsewhere. 
Thus, the system must avoid discarding regions through an early top-$K$ filter, while still preventing the frontier model from reading the entire context.

\subsection{Overview}
\label{sec:overview}

\begin{figure*}[t]
    \centering
    \includegraphics[width=0.98\linewidth]{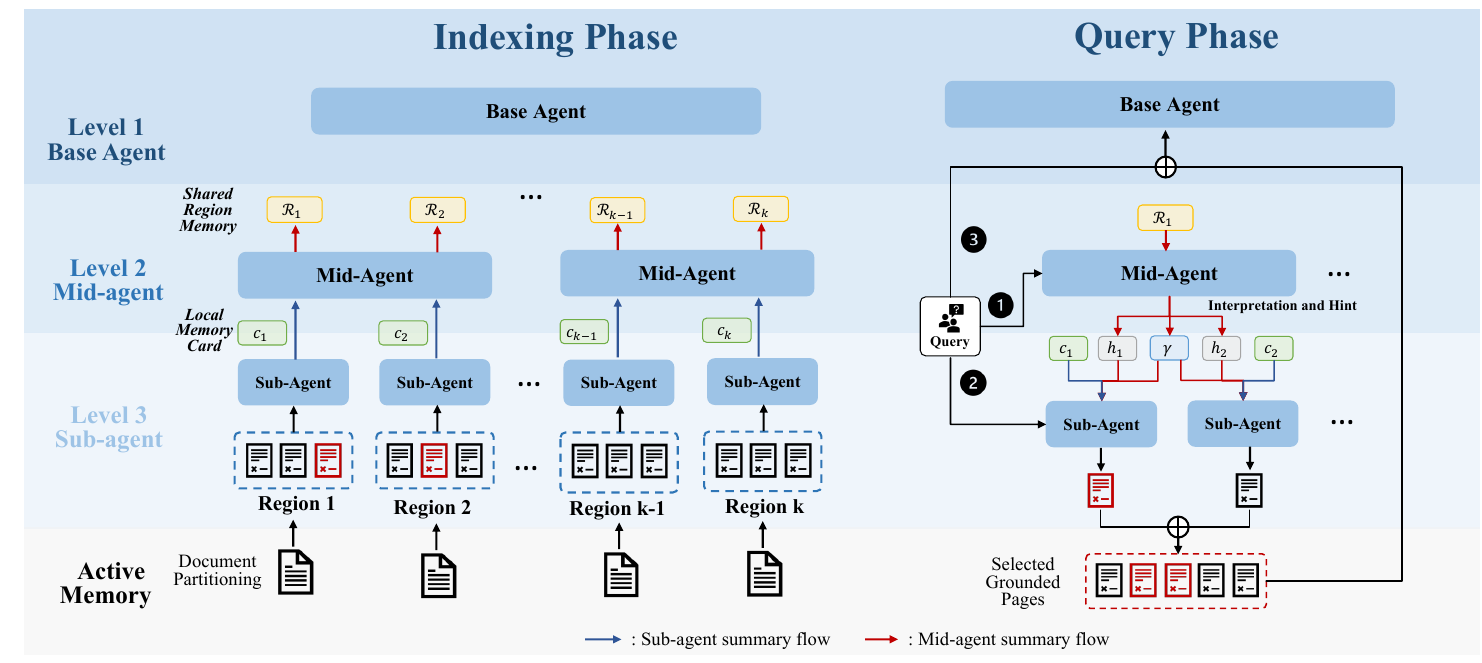}
    \vspace{-6pt}
    \caption{
Overview of \textsc{HMARS}. 
\textbf{Indexing phase:} The document is partitioned into local page-level regions; sub-agents build local memory cards, which mid-agents aggregate into shared regional memory. 
\textbf{Query phase:} Mid-agents broadcast query interpretations and sub-agent-specific hints; all sub-agents select relevant pages in parallel, and the base agent synthesizes the final answer from the ordered grounded evidence.
}
    \vspace{-15pt}
    \label{fig:overview}
\end{figure*}

Given a document and a query, \textsc{HMARS} follows the two-phase workflow in Figure~\ref{fig:overview}. 
In the offline \textit{indexing phase}, the document is tokenized and partitioned into local regions, each managed by a sub-agent with page-level access. 
Sub-agents are grouped under mid-agents, which aggregate local memory cards into shared regional memory. 
In the online \textit{query phase}, the query is routed through the hierarchy to identify relevant grounded pages, which are assembled into the evidence context for final answer synthesis by the base agent.

A key design principle is \emph{asymmetric model allocation}.
We instantiate sub-agents with a lightweight model $\mathcal{L}_{\text{sub}}$, mid-agents with a stronger coordination model $\mathcal{L}_{\text{mid}}$, and the base agent with a frontier model $\mathcal{L}_{\text{base}}$:
\begin{equation}
    \mathcal{L}_{\text{sub}} \ll \mathcal{L}_{\text{mid}} \ll \mathcal{L}_{\text{base}} .
\end{equation}
This separates local memory management, regional coordination, and final reasoning: small models manage local pages, mid-level models share regional context, and the frontier model reasons only over retrieved evidence rather than the full document.

\begin{promptbox}[label={prompt:local-memory-card}]{Local Memory Card Creation}

You are a summarization agent for a long document section.

Below are the pages (numbered) in your section. Write a concise summary
(3--6 sentences) that covers the main topics, entities, events, and any
distinctive facts/numbers mentioned. The summary will be read by a higher-
level agent that routes queries, so be specific about what THIS section
contains versus general background.

\promptfield{Pages:}

\placeholder{pages\_text}

Return ONLY the summary text, no prose preamble.

\end{promptbox}

\paragraph{Indexing phase.}
The document is first partitioned into bounded local regions.
Each region is assigned to a sub-agent $s$.
Within each sub-agent, the assigned tokens are further divided into fixed-size pages
$\{p_{1}, \ldots, p_{n}\}$.
Each page stores its original token offsets, enabling selected pages to be reconstructed in document order.

Each sub-agent then creates a \emph{local memory card}
\begin{equation}
    c = (\text{summary}, \text{topics}, \text{entities}, \text{page\_index}),
\end{equation}
with Prompt~\ref{prompt:local-memory-card},
where $\text{summary}$ describes the local region, $\text{topics}$ and $\text{entities}$ expose salient content, and $\text{page\_index}$ preserves page-level grounding.
The memory card is not used as final evidence; rather, it exposes what the sub-agent knows to its mid-agent.

Sub-agents are grouped under mid-agents.
A mid-agent $m_j$ receives the memory cards $\{c^j_1,...,c_k^j\}$ from its assigned sub-agent group $\mathcal{S}_j=\{s^j_1,...,s_k^j\}$ and constructs a \emph{shared regional memory} with Prompt~\ref{prompt:region-summary}: 
\begin{align}
    \mathcal{R}_j &= m_j(c^j_1,...,c_k^j)\\
    &= (u_j, \text{submap}_j, \text{crossref}_j),
\end{align}
where $u_j$ is a region-level summary, $\text{submap}_j$ maps each sub-agent to its main topics and entities, and $\text{crossref}_j$ records possible cross-sub-agent dependencies such as shared entities, related events, temporal updates, or repeated constraints.
This shared regional memory is the mechanism by which otherwise isolated sub-agents exchange information through the mid-agent.

\begin{promptbox}[label={prompt:region-summary}]{Region Summary}

You manage \placeholder{n} sub-sections of a long document.
Below are the summaries produced by each sub-section.

\placeholder{sub\_summaries}

Write a region-level summary in two parts, as valid JSON:

\placeholder{Json\_Format}

Each sub\_map entry lists the primary topics/entities in that sub-section
(3--6 topics, 3--8 entities). Return ONLY the JSON object.

\end{promptbox}

\paragraph{Query phase.}
Given a query $q$, each mid-agent first interprets the query against its shared regional memory.
Instead of producing a hard routing decision, the mid-agent produces a region-level query interpretation $\gamma_j$ and sub-agent-specific hints with Prompt~\ref{prompt:broadcast} in Appendix~\ref{app:prompt}:
\begin{equation}
    (\gamma_j, \{h_i^j\}_{s_i^j \in \mathcal{S}_j})
    =
    \mathcal{L}_{\text{mid}}
    \big(
    \text{BroadcastPrompt}(q, \mathcal{R}_j)
    \big).
\end{equation}
The interpretation $\gamma_j$ identifies how the query may relate to the region, including relevant entities, constraints, temporal cues, and possible cross-sub-agent dependencies.
Each hint $h_i^j$ tells sub-agent $s_i^j$ what to look for in its own local pages.
For sub-agents that are not strongly implicated by the regional memory, the mid-agent still provides a default query interpretation rather than excluding them from participation.

Then, all sub-agents perform local page selection in parallel.
Each sub-agent receives the query, its own local memory card, and the mid-agent broadcast with Prompt \ref{prompt:select} in Appendix~\ref{app:prompt}:
\begin{equation}
    \mathcal{L}_{\text{sub}}
    \big(
    \text{SelectPrompt}(
    q, c_i^j,
    \underbrace{\gamma_{j}, h_i^j}_{\text{from } m_j},
    \underbrace{p_{1}^i, \ldots, p_{n_i}^i}_{\text{pages for } s_i}
    )
    \big),
\end{equation}
The output is a set of page IDs with optional confidence scores and rationales.
Because every sub-agent performs this local assessment, no document region is removed by an early top-$K$ filter.
At the same time, selection is not independent across sub-agents: each local decision is conditioned on the shared regional interpretation produced by its mid-agent.

Finally, selected pages are deduplicated, sorted by global token offset, and mapped back to their original text spans.
The base agent receives the resulting grounded context $\mathcal{C}(q)$ and generates the answer with Prompt~\ref{prompt:answer} in Appendix~\ref{app:prompt}:
\begin{equation}
    a =
    \mathcal{L}_{\text{base}}
    \big(
    \text{AnswerPrompt}(q, \mathcal{C}(q))
    \big).
\end{equation}

This hierarchy separates three roles: sub-agents manage local page memory, mid-agents provide shared regional context and query-specific information sharing, and the base agent performs final reasoning over grounded evidence.

\section{Experiments}
\label{sec:experiments}

We evaluate \textsc{HMARS} on two complementary benchmarks designed to expose the structural limitations of top-$K$ retrieval: a long-document understanding benchmark requiring cross-chunk synthesis, and a multi-turn memory benchmark requiring integration of scattered conversational facts.
\subsection{Benchmarks}
\label{sec:benchmarks}

We evaluate \textsc{HMARS} on two diagnostic benchmarks designed to isolate the failure modes of passive top-$K$ access: \emph{evidence breadth}, where answers require integrating many scattered pieces of evidence, and \emph{context-dependent relevance}, where a passage becomes useful only under information stated elsewhere. 
The first benchmark targets long-document reasoning over distributed evidence; the second targets cumulative memory over multi-turn interactions. 
Dataset construction details, quality control, and examples are provided in Appendix~\ref{app:dataset}.

\subsubsection{Long-Document Evidence-Breadth Benchmark}
\label{sec:bench_understanding}

We derive a long-document benchmark from LongBench-v2, selecting 29 documents of 20K--120K tokens across legal, scientific, literary, and game-theory domains. 
For each document, we generate and filter questions whose answers require evidence from multiple distant chunks rather than a single localized span. 
Each retained question must require at least four supporting chunks and have embedding-based RAG recall at most 0.5 when retrieving $K$ chunks, where $K$ equals the number of supporting chunks.

The final benchmark contains 195 questions, with an average of 8.6 supporting chunks per question and mean RAG recall of 0.338. 
This makes the benchmark diagnostic of cases where top-$K$ retrieval fails to expose the full evidence set. 
Table~\ref{tab:understanding_stats} summarizes the reasoning-type distribution.

\begin{table}[t]
\centering
\small
\caption{Long-document benchmark statistics. All questions require evidence from at least four chunks and have RAG recall $\leq 0.5$.}
\begin{tabular}{lcc}
\toprule
\textbf{Reasoning Type} & \textbf{Count} & \textbf{Avg. Supp. Chunks} \\
\midrule
Synthesis & 73 & 8.9 \\
Comparison & 47 & 8.4 \\
Causal & 28 & 8.7 \\
Cross-reference & 28 & 8.2 \\
Inference & 19 & 8.5 \\
\midrule
\textbf{Total} & \textbf{195} & \textbf{8.6} \\
\bottomrule
\end{tabular}
\label{tab:understanding_stats}
\end{table}

\subsubsection{Cumulative Interaction Memory Benchmark}
\label{sec:bench_multiturn}

To evaluate context-dependent relevance, we construct a multi-turn memory benchmark inspired by RealMem. 
It contains 120 dialogues across 30 realistic scenarios, including travel planning, apartment search, debugging, and task coordination. 
Each dialogue contains 24--49 turns and 5--9 key facts scattered throughout the interaction, such as names, numbers, dates, preferences, constraints, and user updates.

Each sample ends with a question that requires recalling and integrating facts from across the conversation. 
The benchmark covers five memory subtypes: requirement accumulation, progressive updates, long-range dependency, cross-turn integration, and preference tracking. 
Unlike standard retrieval tasks, relevant turns often have low lexical overlap with the final question and become important only under later context.
\subsection{Experimental Setup}
\label{sec:setup}

\paragraph{Models.}
\textsc{HMARS} uses an asymmetric model hierarchy: Qwen3-0.6B \citep{yang2025qwen3technicalreport} builds sub-agent local memory cards, Qwen3-8B performs mid-agent coordination and local page selection, and GPT-5.1 \citep{singh2026openaigpt5card} performs final answer synthesis.
Although each sub-agent owns a local memory region, its query-time page selection is executed by the Qwen3-8B selection model conditioned on the sub-agent memory card and mid-agent hint.
The Qwen models are served with vLLM. Unless otherwise stated, all inference uses temperature 0.1. Additional decoding parameters and prompts are provided in Appendix~\ref{app:implementation}.

\begin{table*}[h]
    \centering
    \small
        \caption{Main comparison on the long-document understanding and multi-turn memory benchmarks (accuracy in \%). RAG is evaluated with both 1K and 4K retrieval chunks and multiple retrieval budgets. Dense Rerank retrieves $N$ candidates with the same embedder, scores them with Qwen3-8B, and passes the top $K$ chunks to the answer model. GraphRAG uses local search over an LLM-extracted entity graph with community summaries. Chain-of-Agents and MemAgent represent agentic long-context processing and sequential memory baselines. \textsc{HMARS} uses page size $P\!=\!1024$.}
    \setlength{\tabcolsep}{4pt}
    \begin{tabular}{lcccccccc}
    \toprule
    \textbf{Method} & \multicolumn{5}{c}{\textbf{Understanding (195)}} & \textbf{Multi-Turn} & \textbf{Overall} \\
    \cmidrule(lr){2-6} \cmidrule(lr){7-7} \cmidrule(lr){8-8}
     & Syn.\,(73) & Cmp.\,(47) & Cau.\,(28) & Xref.\,(28) & Inf.\,(19) & Acc.\,(120) & Acc.\ \\
    \midrule
    RAG ($K\!=\!10$, 1K)              & 65.8 & 53.2 & 60.7 & 71.4 & 68.4 & 60.8 & 62.2 \\
    RAG ($K\!=\!20$, 1K)              & 82.2 & 66.0 & 60.7 & 82.1 & \textbf{94.7} & 81.7 & 78.4 \\
    RAG ($K\!=\!50$, 1K)              & 75.4 & 57.5 & 68.0 & 75.1 & 79.0 & 86.7 & 76.5 \\
    RAG ($K\!=\!3$, 4K)               & 56.2 & 29.8 & 42.9 & 53.6 & 47.4 & 87.5 & 62.2 \\
    RAG ($K\!=\!5$, 4K)               & 56.1 & 29.8 & 53.5 & 57.0 & 57.8 & 81.7 & 61.9 \\
    Full-Context LLM                  & 74.0 & 59.6 & 53.6 & 75.0 & 84.2 & 80.0 & 73.0 \\
    Dense Rerank ($N\!=\!50$)         & 76.7 & 63.8 & 67.9 & 71.4 & 84.2 & 84.2 & 76.8 \\
    Dense Rerank ($N\!=\!100$)        & 78.1 & 66.0 & 64.3 & 78.6 & 89.5 & 89.2 & 80.0 \\
    GraphRAG~\citep{edge2024lgraphrag}  & 64.4 & 57.4 & 64.3 & 67.9 & 57.9 & 84.0 & 70.7 \\
    CoA-7B~\citep{zhang2024coa}       & 5.5 & 8.5 & 7.1 & 10.8 & 0.0 & 40.0 & 19.6 \\
    MemAgent-14B~\citep{yu2025memagent} & 2.6 & 2.1 & 7.2 & 3.6 & 0.0 & 65.8 & 27.0 \\
    \midrule
    \textsc{HMARS} (ours)             & \textbf{83.0} & \textbf{66.9} & \textbf{70.8} & \textbf{91.7} & 92.0 & \textbf{90.8} & \textbf{83.8} \\

    \bottomrule
    \end{tabular}

    \label{tab:main_results}
    \end{table*}

\paragraph{Baselines.}
We compare \textsc{HMARS} against five families of baselines. 
First, we use dense RAG with Qwen3-Embedding-0.6B and the same frontier answer model as \textsc{HMARS}, varying both retrieval budget and chunk size. 
Second, we evaluate Dense Rerank, which retrieves a large candidate pool and reranks candidates with Qwen3-8B before answer synthesis. 
Third, we include a Full-Context LLM baseline that feeds the entire document, up to the model context limit, directly to the frontier model. 
Fourth, we compare against GraphRAG, which performs local search over an LLM-extracted entity graph. 
Finally, we include agentic long-context baselines: Chain-of-Agents \citep{zhang2024coa}, which sequentially passes compressed communication units across chunk-level workers, and MemAgent \citep{yu2025memagent}, which maintains a sequential memory state while reading long inputs. 
Implementation details are provided in Appendix~\ref{app:implementation}.

\paragraph{Evaluation.}
Following LongBench-v2, we use GPT-based judging: for each question, gold answer, and model prediction, the judge assigns a binary correctness score with a brief justification. 
We report accuracy as the primary metric. 
For efficiency, we additionally report retrieved prompt tokens, and estimated FLOPs per query.

\subsection{Main Results}
\label{sec:main_results}

Table~\ref{tab:main_results} compares \textsc{HMARS} with dense retrieval, reranking, full-context prompting, graph-based retrieval, and agentic long-context memory baselines. 
\textsc{HMARS} achieves the best overall performance and obtains the highest accuracy on the multi-turn benchmark. 
On long-document understanding, it is also the strongest method in most reasoning categories, including synthesis, comparison, causal reasoning, and cross-reference. 
The only category where it does not lead is inference, where a high-budget RAG setting performs best. 
This pattern supports our central claim: when questions require broad evidence coverage or context-dependent memory access, organizing the input as managed local memory is more effective than treating it as a flat retrieval corpus.

The baselines show that the gains are not simply due to retrieving more text. 
Increasing RAG's retrieval budget helps in some cases, but performance remains sensitive to both $K$ and chunk size: coarse 4K chunks hurt long-document understanding, while very large top-$K$ contexts introduce noise. 
Dense Rerank is the strongest non-\textsc{HMARS} baseline, confirming the value of small-model relevance scoring, but it still ranks a flat candidate pool without local memory ownership or mid-level coordination. 
Full-context prompting also underperforms \textsc{HMARS}, indicating that exposing all tokens to a frontier model is not a substitute for structured memory access. 
Chain-of-Agents and MemAgent underperform in our diagnostic setting, suggesting that sequential compression can lose fine-grained evidence when answers require broad, non-adjacent support; we discuss this failure mode in Appendix~\ref{memagent_poor}. 
Together, these results suggest that \textsc{HMARS}'s advantage comes from a hierarchical memory system: local regions assess their own relevance, mid-agents share regional context, and the frontier model reasons over grounded pages.

\subsection{Evidence Coverage Analysis}
\label{sec:evidence_coverage}

To verify that \textsc{HMARS}'s gains come from improved evidence access rather than final-answer prompting, we measure supporting-chunk recall on the long-document benchmark. 
For each question, we compute the fraction of annotated supporting chunks covered by the evidence passed to the answer model. 
This isolates the memory-access stage from the final synthesis stage.

\paragraph{HMARS improves evidence access.}
Table~\ref{tab:evidence_coverage} shows that \textsc{HMARS} covers nearly all annotated supporting chunks before answer generation, while RAG retrieves only a partial evidence set even when its retrieval budget or chunk size is increased. 
This explains the main-result trend: \textsc{HMARS} is not simply benefiting from a different answer prompt, but from a better memory-access mechanism. 
By allowing every local region to perform query-conditioned assessment, \textsc{HMARS} avoids the early evidence loss caused by top-$K$ filtering.

\begin{table}[t]
\centering
\small
\caption{Evidence coverage comparison on the long-document benchmark. }
\resizebox{0.48\textwidth}{!}{%
\begin{tabular}{lccc}
\toprule
\textbf{Method} & \textbf{Supp.\ Recall} & \textbf{Prompt Tok.} & \textbf{FLOPs/query} \\
\midrule
RAG ($K\!=\!5$, 4K)   & 0.571 & $\sim$20K & $\sim$$8.6\!\times\!10^{16}$ \\
RAG ($K\!=\!10$, 1K)  & 0.331 & $\sim$11K & $\sim$$4.4\!\times\!10^{16}$ \\
RAG ($K\!=\!20$, 1K)  & 0.572 & $\sim$20.4K & $\sim$$8.4\!\times\!10^{16}$ \\
\textsc{HMARS}        & \textbf{0.994} & {$\sim$13K} & $\sim$$5.2\!\times\!10^{16}$ \\
\bottomrule
\end{tabular}
}

\label{tab:evidence_coverage}
\end{table}

\paragraph{HMARS improves the recall--cost tradeoff.}
The coverage gain does not come from sending the entire document to the frontier model. 
Compared with high-budget RAG variants, \textsc{HMARS} achieves much higher supporting-chunk recall while passing a smaller and more focused context to the answer model. 
This is the intended role of the hierarchy: lightweight sub-agents provide full-context consideration, and mid-agent scoring compresses the resulting evidence set into grounded pages for final synthesis. 
In contrast, increasing RAG's $K$ or chunk size exposes more text but still misses substantial evidence and eventually introduces noise. 
Thus, \textsc{HMARS} shifts computation from frontier-model reading to structured memory management, yielding a stronger recall--cost operating point.

\subsection{Ablation on Coverage and Coordination}
\label{sec:coverage_coordination}

\begin{figure}[t]
    \centering
    \includegraphics[width=0.95\linewidth]{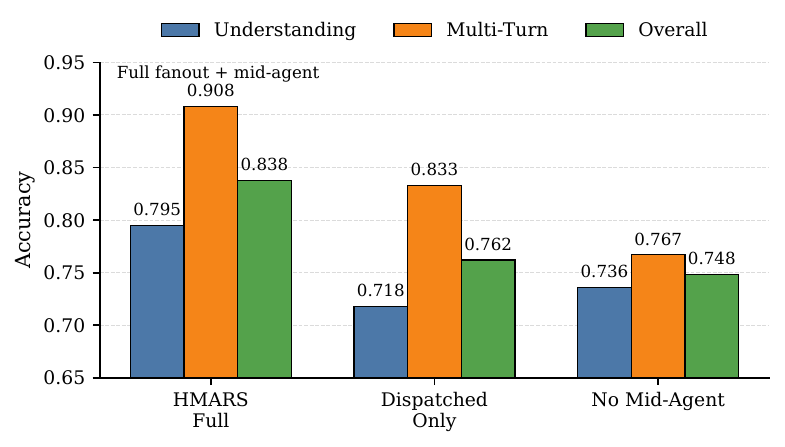}
    \vspace{-6pt}
    \caption{Architectural ablations for \textsc{HMARS}.}
    \vspace{-10pt}
    \label{fig:arch_ablation}
\end{figure}

We first test the two architectural mechanisms most central to \textsc{HMARS}: 
(1) whether every local region receives query-conditioned assessment, and 
(2) whether the mid-agent layer is needed for sharing regional context.
Figure~\ref{fig:arch_ablation} compares the anchor system against two variants: \texttt{dispatched}, where only mid-agent-selected sub-agents participate, and \texttt{no-mid}, where the mid-agent layer is removed.

\paragraph{Full-fanout provides coverage.}
Restricting selection to only dispatched sub-agents drops accuracy from 0.838 to 0.762 overall.
The drop is consistent on both benchmarks, showing that summary-level dispatch alone cannot reliably decide which local regions matter.
A region that appears irrelevant in a mid-agent summary may still contain pages that become relevant only after the query is paired with the local content.
This supports the core design of \textsc{HMARS}: every local memory region should receive query-conditioned assessment before evidence is discarded.

\paragraph{Mid-agents provide shared regional context.}
Removing the mid-agent layer drops overall accuracy from 0.838 to 0.748, with the largest degradation on multi-turn memory.
This indicates that sub-agents should not operate as isolated local readers.
The mid-agent aggregates sub-agent memories into a shared regional view and broadcasts query-specific guidance, helping local agents interpret their pages under broader context.
Thus, full fanout and mid-agent coordination are complementary: full fanout ensures coverage, while mid-agents improve how local memories are interpreted.

\vspace{-2pt}

\subsection{Ablation on Memory Granularity}
\label{sec:granularity_ablation}

\begin{figure}[t]
    \centering
    \includegraphics[width=0.95\linewidth]{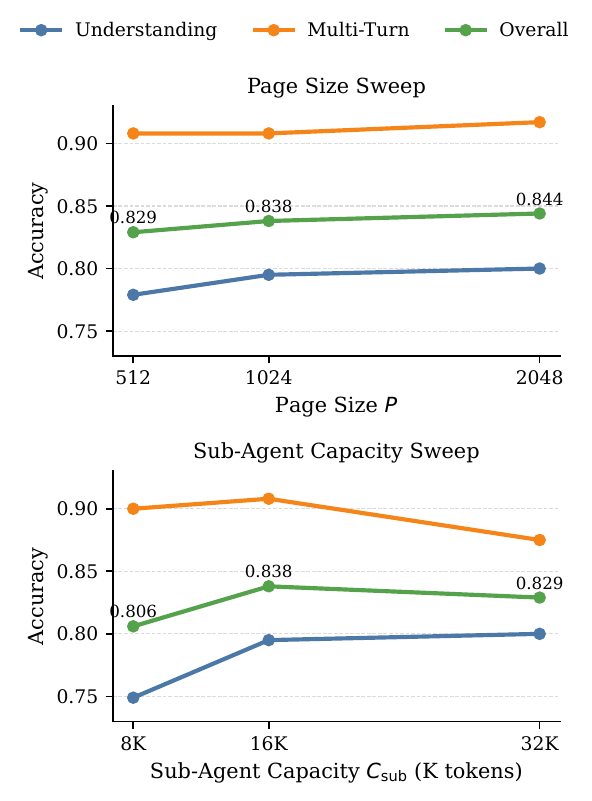}
    \vspace{-6pt}
    \caption{
    Granularity ablations for \textsc{HMARS}. 
    Page size $P$ controls the evidence unit returned to the base model, while sub-agent capacity $C_{\text{sub}}$ controls the size of each local memory region. 
    }
    \vspace{-10pt}
    \label{fig:granularity_ablation}
\end{figure}

We next study how sensitive \textsc{HMARS} is to memory granularity: the page size $P$, which controls the evidence unit returned to base model, and the sub-agent capacity $C_{\text{sub}}$, which controls how much context each sub-agent owns.
Figure~\ref{fig:granularity_ablation} shows both sweeps.

\paragraph{Page size is not the main driver.}
Changing $P$ across $\{512,1024,2048\}$ yields similar performance, with all overall scores within a narrow range.
Smaller pages provide finer-grained grounded evidence, while larger pages preserve more local context and reduce page-scoring overhead.
The small variation suggests \textsc{HMARS} does not depend on a finely tuned page size; its gains mainly come from the memory-management structure rather than a particular evidence granularity.

\paragraph{Sub-agent capacity controls the memory partition.}
Changing $C_{\text{sub}}$ has a clearer effect.
A smaller capacity fragments the document into too many local memories, while a larger capacity makes each local memory coarser and hurts multi-turn memory.
The default $16K$ capacity provides a balanced partition, preserving enough local context within each sub-agent while keeping regional coordination manageable.

\vspace{-2pt}

\vspace{-2pt}

\subsection{Ablation on Model Allocation}
\label{sec:model_allocation}

Finally, we test whether local memory management requires a larger sub-agent model, results are shown in Table \ref{tab:model_ablation}.
Replacing the 0.6B sub-agent model with the 8B model yields only a small change on long-document understanding and reduces multi-turn accuracy.
Overall accuracy changes from 0.838 to 0.832, suggesting that scaling the sub-agent model does not provide a reliable gain.

\begin{table}[h]
\centering
\small
\caption{Sub-agent model size ablation.}
\begin{tabular}{lccc}
\toprule
\textbf{Sub-agent Model} & \textbf{UND} & \textbf{MT} & \textbf{Overall} \\
\midrule
Qwen3-0.6B & 0.795 & \textbf{0.908} & \textbf{0.838} \\
Qwen3-8B   & \textbf{0.800} & 0.883 & 0.832 \\
\bottomrule
\end{tabular}
 
\label{tab:model_ablation}
\end{table}

This supports the asymmetric allocation used in \textsc{HMARS}: lightweight models are sufficient for local memory summarization and page-level management, while stronger models are better reserved for mid-level coordination and final synthesis.

\vspace{-2pt}

\vspace{-2pt}

\subsection{Efficiency Analysis}
\label{sec:efficiency}

\begin{figure}[t]
    \centering
    \includegraphics[width=0.9\linewidth]{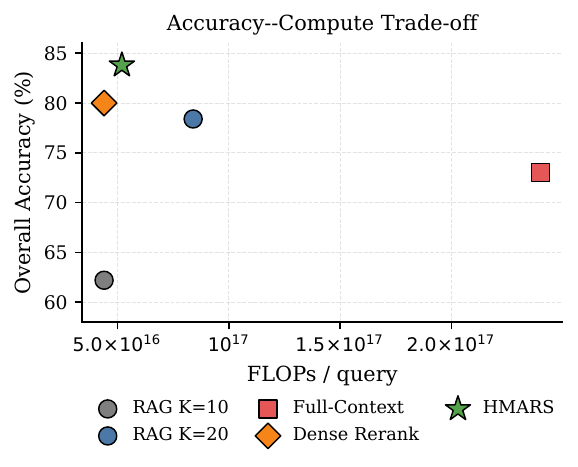}
    \vspace{-6pt}
    \caption{Accuracy--compute trade-off on the full evaluation set. }
   
    \label{fig:efficiency_tradeoff}
\end{figure}

Figure~\ref{fig:efficiency_tradeoff} compares representative methods by overall accuracy and estimated per-query FLOPs. 
\textsc{HMARS} achieves the best overall accuracy with roughly $5.2\times10^{16}$ FLOPs per query, only modestly above RAG ($K\!=\!10$) and Dense Rerank ($4.4\times10^{16}$), but well below high-budget RAG and Full-Context prompting. 
This efficiency comes from asymmetric model allocation: local and mid-level models perform memory access over the full input, while the frontier model reasons only over selected grounded pages.

The comparison shows that more compute does not necessarily lead to better long-context reasoning. 
Full-Context prompting uses the largest budget ($2.4\times10^{17}$ FLOPs) but underperforms \textsc{HMARS}; RAG with a larger retrieval budget reaches $8.4\times10^{16}$ FLOPs while still trailing. 
Dense Rerank is efficient and strong, but remains a flat candidate-ranking pipeline. 
Overall, \textsc{HMARS} offers a favorable accuracy--compute trade-off by shifting work from frontier-model reading to structured memory access.
\vspace{-2pt}

\section{Conclusion} \label{sec:conclusion} \vspace{-2pt} We introduced \textsc{HMARS}, a hierarchical multi-agent memory system for long-context reasoning over documents and interaction histories. Rather than treating long inputs as flat retrieval corpora, \textsc{HMARS} organizes them into local memory regions, uses mid-agents to coordinate query-specific regional context, and reserves the frontier model for reasoning over grounded evidence. Across diagnostic benchmarks targeting evidence breadth and context-dependent relevance, \textsc{HMARS} improves over retrieval, reranking, full-context, graph-based, and agentic baselines. These results suggest that long-context reasoning should be framed not only as retrieving more chunks or extending context windows, but as building memory systems that access, coordinate, and ground evidence across the input.

\clearpage
\section*{Limitations}
\label{sec:limitations}

\textsc{HMARS} provides full-context consideration, but not guaranteed evidence recall: local agents can still miss relevant pages or select noisy evidence. 
The current implementation also relies on prompted LLM agents rather than learned memory policies, making performance sensitive to prompts, model choice, and decoding settings. 
While the asymmetric hierarchy reduces frontier-model context cost, it adds local and mid-level model calls, so practical efficiency depends on parallel serving. 
Empirically, the benefits are strongest for cumulative interaction memory; on long-document understanding, stronger RAG settings with larger $K$ or chunk size can sometimes outperform \textsc{HMARS}. 
Finally, our diagnostic benchmarks are designed to stress evidence breadth and context-dependent relevance, and should be complemented with broader natural workloads and additional agentic-memory baselines in future work.

\bibliography{custom}

\clearpage

\appendix

\section{Benchmark Construction Details}
\label{app:dataset}

\subsection{Long-Document Benchmark Construction}\label{app:long-doc}

This appendix provides additional details on the construction of the
long-document understanding benchmark introduced in
Section~\ref{sec:bench_understanding}.

\paragraph{Source document selection.}
We derive the long-document benchmark from LongBench-v2, selecting 29 English documents between 20K and 120K tokens across legal texts, scientific papers, literary criticism, and game-theory transcripts.

\paragraph{Question generation.}
For each document, we prompt GPT-5.1 to generate candidate questions
in the five categories defined in Section~\ref{sec:benchmarks}
(Synthesis, Comparison, Causal, Cross-reference, Inference).  The
generation prompt instructs the model to:
\begin{enumerate}[nosep]
  \item Identify 3--5 non-trivial facts spread across different sections.
  \item Formulate a question whose correct answer requires integrating
        all identified facts.
  \item Provide a gold-standard answer and list the minimal set of
        supporting passages (by section or paragraph index).
\end{enumerate}
We generate 10--15 candidate questions per document to allow for
attrition during quality control.

\paragraph{Supporting chunk annotation.}
We segment each document into 1,024-token chunks and map supporting passages to chunk indices. A question is retained only if it requires at least four supporting chunks.
\paragraph{Quality control and RAG recall filtering.}
We compute embedding-based RAG recall using the same embedding model as our RAG baseline. For each question, we retrieve \(K\) chunks, where \(K\) equals the number of annotated supporting chunks. We retain only questions with RAG recall \(\leq 0.5\). After filtering, the benchmark contains 195 questions.

\paragraph{Example.}
\begin{itemize}[nosep]
  \item \textbf{Type:} Synthesis
  \item \textbf{Question:} ``Based on Sections 3, 5, and 7 of the report,
        what are the three main factors contributing to the observed
        decline in biodiversity, and how do the authors propose to
        address each?''
  \item \textbf{Supporting chunks:} 12, 34, 58 (out of 72 total chunks)
  \item \textbf{Gold answer:} (3-sentence summary linking habitat loss,
        climate change, and invasive species to the proposed mitigation
        strategies.)
\end{itemize}

\subsection{Cumulative Interaction Benchmark Construction}\label{app:dialogue}

This appendix details the construction of the cumulative multi-turn
dialogue benchmark introduced in Section~\ref{sec:bench_multiturn}.

\paragraph{Scenario design.}
We generate 120 dialogues across 30 realistic scenarios, including travel planning, apartment search, debugging, and task coordination.

\paragraph{Dialogue generation.}
Each dialogue contains 24--49 turns, with 5--9 key facts scattered throughout the interaction. Key facts include names, numbers, dates, preferences, constraints, and user updates.

\paragraph{Key fact placement.}
Key facts are embedded at pre-designated turns (e.g.\ turns 2, 5, 9, 12)
so that the final probe question requires the model to recall and
integrate information from multiple, temporally distant points in the
conversation.  On average, each dialogue contains 5.1 key facts spread
across 4.2 distinct turns.

\paragraph{Subtype definitions.}
The 120 probe questions are evenly divided into three subtypes:
\begin{itemize}[nosep]
  \item \textbf{Fact Recall} (40 questions): Requires retrieving a specific
        fact stated in an earlier turn.
  \item \textbf{Temporal Ordering} (40 questions): Requires determining the
        chronological sequence of events or decisions discussed across turns.
  \item \textbf{Cross-Turn Inference} (40 questions): Requires combining
        facts from $\ge$2 non-adjacent turns to derive a conclusion not
        stated in any single turn.
\end{itemize}

\paragraph{Example.}
\begin{itemize}[nosep]
  \item \textbf{Scenario:} Chronic disease management (Healthcare)
  \item \textbf{Turns:} 12 (average history length: 11.8\,K tokens)
  \item \textbf{Probe (Cross-Turn Inference):} ``Given the patient's
        updated lab results from turn 9 and the medication change
        discussed in turn 4, should the treatment plan be revised?''
  \item \textbf{Key facts:} Medication dosage (turn 4), lab results
        (turn 9), contraindication guideline (turn 2)
\end{itemize}

\section{Implementation Details and Hyperparameters}
\label{app:implementation}

\paragraph{Model hierarchy.}
\textsc{HMARS} uses an asymmetric model hierarchy. 
Sub-agent local memory cards are generated with Qwen3-0.6B, mid-agent coordination and page selection are performed with Qwen3-8B, and final answer synthesis is performed by GPT-5.1 model. 
The Qwen models are served with vLLM. 
Unless otherwise stated, all model calls use temperature $0.1$.
We run all the experiments on NVIDIA H100 GPU.
\paragraph{Document partitioning.}
All splitting is performed with the model tokenizer rather than character length. 
Given a document or interaction history $\mathcal{D}$, we partition it into local memory regions of at most
$C_{\mathrm{sub}}=16{,}384$ tokens. 
Each local region is assigned to one sub-agent and further divided into pages of size $P=1024$ tokens. 
Each page stores its global token offsets, so selected pages can be mapped back to the original document and ordered before final synthesis.

\paragraph{Agent grouping.}
Sub-agents are grouped under mid-agents with maximum group size $G=10$. 
If the document contains $M$ sub-agents, the system creates
$K=\lceil M/G\rceil$ mid-agents. 
The final group may contain fewer than $G$ sub-agents. 
Each mid-agent constructs a shared regional memory from the local memory cards of its assigned sub-agents.

\paragraph{Query-time configuration.}
At query time, \textsc{HMARS} uses full-fanout page selection: every sub-agent receives the query and participates in local page assessment. 
Mid-agents do not serve as hard gates; instead, they produce a region-level query interpretation and sub-agent-specific hints. 
Sub-agents then select relevant page IDs conditioned on the query, their local memory card, and the mid-agent hint. 
Selected pages are deduplicated, sorted by global token offset, and passed to the frontier model as the final evidence context.

\paragraph{Baseline settings.}
For dense RAG, we use Qwen3-Embedding-0.6B with cosine similarity and the same frontier answer model as \textsc{HMARS}. 
We evaluate both 1K-token and 4K-token retrieval chunks with multiple retrieval budgets. 
For Dense Rerank, we first retrieve the top-$N$ candidates using the same dense retriever, then score candidates with Qwen3-8B and pass the top-ranked chunks to the frontier model. 
Full-Context LLM directly feeds the document to the frontier model up to the model context limit. 
GraphRAG, Chain-of-Agents, and MemAgent follow their standard or released inference protocols when applicable.

\paragraph{Evaluation.}
We use GPT-based judging following LongBench-v2. 
For each question, gold answer, and predicted answer, the judge assigns a binary correctness score with a short justification. 
Accuracy is the primary metric. 
For efficiency analysis, we report latency, frontier prompt tokens, and estimated per-query FLOPs. 
Prompt tokens refer to the text passed to the frontier answer model, not the full text processed by lightweight sub-agents or mid-agents.

\section{Why Full-Context Consideration Helps}
\label{app:full_fanout}

Consider a query $q$ whose answer requires evidence from a set of supporting pages or chunks
$\mathcal{E} = \{e_1,\ldots,e_R\}$ distributed across the input. In a standard RAG pipeline, each chunk $c_i$ is scored independently by a similarity function $\mathrm{sim}(q,c_i)$, and only the top-$K$ chunks are passed to the answer model. The evidence recall is

\begin{equation}
    \mathrm{Recall}_{\mathrm{RAG}}
    =
    \frac{|\{e \in \mathcal{E}: e \in \mathrm{top}\text{-}K\}|}
    {|\mathcal{E}|}.
\end{equation}

This mechanism can fail even when the evidence is present in the input. A passage may use vocabulary different from the query, or its relevance may depend on information stated elsewhere. For example, a question about ``housing constraints'' may require an earlier turn mentioning a ``\$2,500 monthly budget,'' despite limited lexical overlap.

\textsc{HMARS} removes this early exclusion bottleneck by giving every local memory region an opportunity to perform query-conditioned assessment. This should not be interpreted as guaranteeing perfect evidence recall: local agents may still fail to select a relevant page. Rather, the system guarantees \emph{consideration coverage}: no region is discarded before a local agent evaluates its pages under the query and the mid-agent's regional guidance.

This design is most useful in two settings. First, in \emph{evidence-breadth} questions, the answer depends on many non-adjacent regions, so a small top-$K$ set is unlikely to contain all supporting evidence. Second, in \emph{context-dependent relevance} questions, whether a passage matters can only be determined after the query is interpreted against broader memory state. These are precisely the settings targeted by our diagnostic benchmarks.

\section{Computational Complexity}
\label{app:complexity}

Let $N$ denote the document length in tokens, $P$ the page size, $C_{\mathrm{sub}}$ the maximum token capacity of each sub-agent, and $G$ the maximum number of sub-agents assigned to one mid-agent. The number of sub-agents and mid-agents are

\begin{equation}
    M = \left\lceil \frac{N}{C_{\mathrm{sub}}} \right\rceil,
    \qquad
    K = \left\lceil \frac{M}{G} \right\rceil.
\end{equation}

\paragraph{Indexing cost.}
During indexing, each sub-agent summarizes its assigned region into a local memory card, and each mid-agent aggregates a group of local memory cards into shared regional memory. If $C_{\mathrm{sum}}$ and $C_{\mathrm{reg}}$ denote the average token costs of the local and regional summarization calls, then the indexing cost scales as

\begin{equation}
    \mathrm{FLOPs}_{\mathrm{index}}
    \approx
    2|\theta_{\mathrm{sub}}| M C_{\mathrm{sum}}
    +
    2|\theta_{\mathrm{mid}}| K C_{\mathrm{reg}}.
\end{equation}

This cost is amortized across all queries over the same document or interaction history.

\paragraph{Query cost.}
At query time, each mid-agent produces a regional query interpretation and sub-agent-specific hints. Then local page selection is performed for each sub-agent, followed by one frontier-model synthesis call. Let $\theta_{\mathrm{sel}}$ denote the model used for page selection; in our implementation, this is the Qwen3-8B model used for mid-level coordination. The query cost scales as

\begin{equation}
\begin{split}
\mathrm{FLOPs}_{\mathrm{query}}
\approx &\ 2|\theta_{\mathrm{mid}}|K C_{\mathrm{broadcast}} \\
& + 2|\theta_{\mathrm{sel}}|M C_{\mathrm{select}} \\
& + 2|\theta_{\mathrm{base}}| C_{\mathrm{ans}}.
\end{split}
\end{equation}

The main efficiency advantage comes from asymmetric model allocation. Full-context consideration is performed by smaller local and mid-level models, while the frontier model only receives the selected grounded evidence.

\paragraph{Parallelism.}
The mid-agent calls and sub-agent selection calls can be executed concurrently within each phase. In practice, latency is dominated by the slowest local or mid-level call plus the final frontier-model synthesis call, rather than by the sum of all local calls.

\section{Additional Baseline Notes}
\label{app:additional_baselines}

\paragraph{Dense retrieval with reranking.}
Dense Rerank is the strongest flat retrieval baseline in our experiments. It first retrieves a large candidate set using the same embedding model as RAG, then scores candidates with Qwen3-8B before passing the top-ranked chunks to the frontier answer model. Its strong performance supports one of our main design intuitions: small-model relevance scoring is a useful primitive. However, it remains a flat candidate-ranking pipeline. It does not assign persistent local memory ownership, does not construct shared regional memory, and does not allow all memory regions to participate in query-conditioned assessment.

\paragraph{RAG chunk size and retrieval budget.}
We evaluate RAG with both 1K and 4K retrieval chunks. These chunk sizes play a different role from \textsc{HMARS} page size. A RAG chunk is both the retrieval unit and the text passed to the answer model, so larger chunks may preserve more local context but also dilute retrieval precision and increase prompt noise. In contrast, a \textsc{HMARS} page is a grounded evidence unit selected within a larger agent-managed memory region. This distinction explains why increasing RAG's chunk size does not consistently improve performance across our benchmarks.

\paragraph{Full-context prompting.}
Full-context prompting gives the frontier model access to more tokens, but it does not provide an explicit memory interface. The model must still locate scattered evidence among distractors. Our results show that exposing the full document to the frontier model is not a reliable upper bound for long-context reasoning.

\paragraph{GraphRAG.}
GraphRAG provides an entity-centric abstraction over documents and performs local search over an extracted graph. This can be useful when questions are primarily entity-centric, but graph/community summaries may lose fine-grained passage-level evidence needed for broad cross-region synthesis.

\paragraph{Agentic long-context baselines.}
Chain-of-Agents and MemAgent represent sequential agentic processing and rolling-memory approaches. They demonstrate that agentic reading can extend beyond a single prompt, but they compress information into communication units or evolving memory states. In our diagnostic benchmarks, this compression can discard fine-grained evidence. \textsc{HMARS} instead preserves grounded page access throughout the hierarchy.

\paragraph{Why CoA and MemAgent underperform in our setting.}
\label{memagent_poor}
CoA-7B and MemAgent-14B represent sequential agentic long-context processing, but their design assumptions differ from the diagnostic setting studied here.
Both methods compress information during a single pass over the input: CoA propagates a communication unit across chunk-level workers, while MemAgent maintains an evolving memory state.
This compression is useful for reducing context length, but can lose fine-grained evidence when answers require multiple non-adjacent supporting passages or when an early passage becomes relevant only after later context is observed.
In representative errors, the required evidence was often present in the input but absent from the information available to the final answer model.
Thus, these results should not be read as showing that CoA or MemAgent are weak methods in general; rather, they illustrate a mismatch between sequential compression-based reading and our evidence-breadth, context-dependent memory setting.
\textsc{HMARS} is designed for this failure mode by preserving grounded page access and allowing all local memory regions to participate in query-conditioned assessment.

\section{Prompt Details}
\label{app:prompt}

\begin{promptbox}[label={prompt:answer}]{AnswerPrompt}
You are an expert reader. Answer the question STRICTLY based on the retrieved passages below. If the answer is not supported by the passages, say so explicitly. Be concise (1--4 sentences) and factual.

Retrieved passages:

\placeholder{context}

Question:

\placeholder{question}

Answer:
\end{promptbox}

\begin{promptbox}[label={prompt:local-memory}]{LocalMemoryCardPrompt}
You are a local memory agent for one section of a long document or interaction history.

Below are the pages in your assigned section. Write a concise memory card that captures the main topics, entities, events, constraints, numbers, and distinctive facts in this section. Be specific about what this section contains, since a mid-level agent will use your memory card to coordinate future queries.

Pages:

\placeholder{pages\_text}

Return ONLY a JSON object with the following fields:

\placeholder{JSON\_form}
\end{promptbox}

\begin{promptbox}[label={prompt:broadcast}]{BroadcastPrompt}
You are a mid-level memory coordinator. You manage several local memory regions.

Your regional summary:

\placeholder{region\_summary}

Your local regions, with topics and entities:

\placeholder{sub\_map\_text}

Query:

\placeholder{query}

Interpret how the query may relate to this region. Then select every local region whose content could plausibly contain useful evidence. Do NOT artificially limit the number of selected regions. For each selected region, write a focused hint telling the local agent what to look for, including relevant entities, numbers, constraints, temporal cues, or sub-topics.

Return ONLY a JSON object of this form:

\placeholder{JSON\_form}
\end{promptbox}

\begin{promptbox}[label={prompt:select}]{SelectPrompt}
You are a local memory agent for one section of a long document or interaction history. Select every page from your section that could plausibly help answer the query. Be greedy: when in doubt, include the page. Do not limit the number of pages you return.

Regional context from your mid-level coordinator:

\placeholder{region\_summary}

What the coordinator believes your section contains:

\hspace{1em}topics: \placeholder{sub\_topics}

\hspace{1em}entities: \placeholder{sub\_entities}

Focused hint from your coordinator:

\placeholder{hint}

Query:

\placeholder{query}

Pages in your section:

\placeholder{pages\_text}

Return ONLY a JSON object of this form:

\placeholder{JSON\_form}
\end{promptbox}

\section{Dataset and Artifact Documentation}
\paragraph{Data sources.}
Our long-document benchmark is derived from LongBench-v2, and our cumulative interaction benchmark is inspired by RealMem. We use these sources only for research evaluation and follow their intended benchmark usage.

\paragraph{Dataset construction.}
We describe the document selection, question generation, filtering criteria, supporting-evidence annotation, dialogue scenario generation, and task subtype definitions in Appendix~\ref{app:dataset}.

\paragraph{Statistics.}
The final benchmark contains 195 long-document questions and 120 multi-turn interaction questions. The long-document benchmark contains 29 documents with lengths between 20K and 120K tokens, and each retained question requires at least four supporting chunks. The multi-turn benchmark contains 120 dialogues across 30 scenarios, with 24--49 turns per dialogue and 5--9 key facts per sample.

\paragraph{PII and offensive content.}
The constructed evaluation data consists of benchmark documents and synthetic task-oriented dialogues. We do not intentionally include personally identifying information or offensive content. For synthetic dialogues, names, dates, constraints, and preferences are fictional.

\section{Licenses and Intended Use}
\paragraph{Licenses.}
Our benchmark construction uses LongBench-v2 and RealMem-style settings as research artifacts. We cite the original creators in the main paper and use the data only for academic evaluation. Users should follow the licenses and usage terms of the original benchmark sources.

\paragraph{Intended use.}
The constructed evaluation sets are intended for evaluating long-context question answering, evidence-breadth reasoning, and multi-turn interaction memory. They are not intended for training deployed assistants, profiling real users, or making decisions about individuals.
\paragraph{Release format.}
We will release the evaluation data in JSONL format, including the input document or dialogue, question, gold answer, supporting evidence metadata, and task subtype when applicable.

\section{Use of AI Assistants}
We also used LLMs to generate candidate questions and synthetic dialogues during benchmark construction. All generated examples were filtered according to the criteria described in Appendix~\ref{app:dataset}.

\section{Human Subjects}
This work does not involve human subjects, recruited participants, or paid annotators. The multi-turn interaction benchmark consists of synthetic task-oriented dialogues generated for evaluation purposes. Therefore, participant instructions, recruitment, payment, consent, and IRB approval are not applicable.

\end{document}